\documentstyle[epsfig,float]{aipproc2}
\begin{document}
\title{Realistic Effective Interactions \\ and Nuclear Structure
Calculations}

\author{A. Covello,$^1$ L. Coraggio,$^1$ A. Gargano,$^1$ N. Itaco,$^1$
and T. T. S. Kuo$^2$}
\address {$^1$Dipartimento di Scienze Fisiche, Universit\`a di Napoli Federico II,
and Istituto Nazionale di Fisica Nucleare, \\
Complesso Universitario di Monte S. Angelo, Via Cintia, 80126 Napoli, Italy \\
$^2$Department of Physics, SUNY, Stony Brook, New York 11794}

%\lefthead{LEFT head}
%\rig436thead{RIGHT head}
\maketitle

\begin{abstract}
We address the two main questions relevant to microscopic nuclear
structure calculations starting from a free $NN$ potential.
These concern the accuracy of these kinds of calculations and the
extent to which they depend on the potential used 
as input. Regarding the first question, we present some results obtained
for nuclei around doubly magic $^{132}$Sn and $^{208}$Pb by making 
use of an effective interaction derived from the Bonn A potential.
Comparison shows that our results are in very good agreement
with the experimental data. As for the second question, we present
the results obtained for the nucleus $^{134}$Te by making use
of four different $NN$ potentials. They indicate
that nuclear structure calculations may help in understanding
the off-shell nature of the $NN$ potential. 
\end{abstract}

\section*{Introduction}
%\[
%\widehat{a} + \widehat{ab} + \widehat{abc} + \widehat{abcd}
%\]
%
%%\show\frak
% 
%\[
%%      {\bf x}^{\bf x} \triangleq z 
%      {\bf x}^{\bf x}\triangleq{z} \tensor{T} \frak{E^E}=\frak{mc}^2
%%      {\bf x}^{\bf x}\triangleq {z} \tensor{T} \frak{E}=\frak{mc}^2
%\]
% 
%\[
%{\Bbb {NQRZ}} \qquad \because \eth\ggg\bigstar \therefore\blacktriangleright\rightsquigarrow \blacksquare
%\]
% 

The shell model is the basic framework for nuclear structure calculations
in terms of nucleons. Since the early 1950s many hundreds of shell-model
calculations have been carried out, most of them being very successful
in describing a variety of nuclear structure phenomena. In any standard
shell-model calculation one has to start by defining a model space,
namely by specifying a set of active single-particle (s.p.) orbits.
The choice of the model space is of course conditioned by the size
of the matrices to be set up and diagonalized. The rapid increase in
computer power and the development of high-quality codes in the last decade
has greatly extended the feasibility of large-scale calculations
\cite{Caur98}.
While these technical improvements add to the practical value of the
shell model, much uncertainty still exists for what concerns the model-space
effective interaction $V_{\rm eff}$. In most of the existing calculations
to date either empirical effective interaction containing several adjustable
parameters have been used or the two-body matrix elements have been
treated as free parameters, this latter approach being limited to
small model spaces. 

This uncertainty in shell model work can only
be removed by taking a more fundamental approach, namely by deriving
the effective interaction from the free nucleon-nucleon $(NN)$
potential. As is well known, the first step in this direction was
taken in the mid 1960s by Kuo and Brown \cite{KB66} who derived
an $s$-$d$ shell effective interaction from the Hamada-Johnston
potential \cite{HJ62}. Since that time there has been substantial
progress towards a microscopic approach to nuclear structure
calculations starting from a free $NN$ potential. 
On the one hand, high-quality $NN$ potentials have been constructed
which reproduce quite accurately all the known $NN$ data. On the
other hand, the many-body methods for calculating the matrix
elements of the effective interaction have been largely improved.
A review of modern $NN$ potentials is given in Ref. \cite{Machl98}
while the main aspects of the derivation of $V_{\rm eff}$ are
discussed in Ref. \cite{Kuo96}. These improvements have brought
about renewed interest in shell-model calculations with
realistic effective interactions. In this context, the two crucial
questions are: i) how accurate is an effective interaction derived
from the $NN$ potential? ii) to which extent can nuclear structure
calculations distinguish between different $NN$ potentials?

Recent calculations for nuclei in the $^{100}$Sn and
$^{132}$Sn regions \cite{Andr96,Andr97,Cov97,Holt97,Suho98,Holt98}
have achieved very good agreement with
experiment indicating the ability of realistic effective
interactions to provide a description of nuclear structure properties
at least as accurate as that provided by traditional, empirical
interactions. To our knowledge, no systematic investigation concerning the
second question has been carried out thus far. The main interest
in trying to answer this question stems from the fact that two
potentials which fit equally well the $NN$ data up to the inelastic
threshold may differ substantially in their off-shell behavior.
Thus, from microscopic nuclear structure calculations we may
learn something about the off-shell properties of the nuclear
potential.  

The main aim  of this paper is to report on some achievements of our 
current work relevant to both the above questions. We shall first
present some results of realistic shell-model calculations for
nuclei having either few protons outside doubly magic $^{132}$Sn or
few neutron holes in doubly magic $^{208}$Pb. They are 
$^{135}$I, $^{136}$Xe and $^{206,205,204}$Pb. In all of these
calculations we have made use of a realistic effective interaction
derived from the Bonn A free $NN$ potential \cite{Machl87}.
Then we shall present the results obtained for the two proton-nucleus 
$^{134}$Te by making use of four different potentials, Paris \cite{Lacom80}, 
Nijmegen93 \cite{Stoks94}, Bonn A and CD Bonn \cite{Machl96}, which are all 
based on the meson theory of nuclear force. 

We shall see that while the former study confirms what was
learned from our previous calculations with the Bonn potential,
the latter indicates a dependence of nuclear structure results
on the kind of potential used as input.

\section*{Outline of calculations}
As already mentioned in
the Introduction, for all the six nuclei
considered in this paper we have employed an effective interaction
derived from the Bonn A potential. For $^{134}$Te we have also
performed calculations employing three other effective
interactions derived from the Paris, Nijmegen93 and CD Bonn 
potential, respectively. These effective interactions were all
obtained using a $G$-matrix folded-diagram
formalism, including renormalizations from both core polarization 
and folded diagrams. For the $N=82$ isotones $^{134}$Te, $^{135}$I and
$^{136}$Xe we have considered $^{132}$Sn as an inert core and let
the valence protons occupy the five single-particle
(s.p.) orbits $0g_{7/2}$, $1d_{5/2}$, 
$2s_{1/2}$, $1d_{3/2}$, and $0h_{11/2}$. For the Pb isotopes, we
have treated neutrons as valence holes with respect to the $^{208}$Pb
closed core and included in the model space the six single-hole (s.h.)
orbits $2p_{1/2}$, $1f_{5/2}$, 
$2p_{3/2}$, $0i_{13/2}$, $1f_{7/2}$, and $0h_{9/2}$. A description
of the derivation of our $V_{\rm eff}$ 
for the $N=82$ isotones and for the Pb isotopes can be found in 
Refs. \cite{Cov98} and \cite{Cor98}, respectively. 
For the shell-model oscillator parameter $\hbar \omega$ we have
used the value 7.88 MeV for the $N=82$ isotones and 6.88 MeV for
the Pb isotopes, as obtained from the relationship
$\hbar \omega= 45A^{- 1/3} - 25A^{-2/3}$ for A= 132 and A= 208,
respectively.

As regards the s.p. energies, for the $N=82$ isotones we have taken
three s.p. spacings from the experimental spectrum of $^{133}$Sb
\cite{Serg86,Sanch98}.
In fact, the $g_{7/2}$, $d_{5/2}$, $d_{3/2}$, and $h_{11/2}$
states can be associated with the ground state and the 0.962,
2.439 and 2.793 MeV excited levels, respectively.
As for the $s_{1/2}$ state, its position has been determined 
by reproducing the experimental energy of the ${1 \over 2}^+$ level 
at 2.15 MeV in $^{137}$Cs. This yields the value 
$\epsilon_{s_{1/2}}$ = 2.8 MeV. 
Regarding the Pb isotopes, the s.h. energies have all been taken
from the experimental spectrum of $^{207}$Pb \cite{Martin93}.
They are (in MeV) $\epsilon_{p_{1/2}}=0$, $\epsilon_{f_{5/2}}=0.570$, 
$\epsilon_{p_{3/2}}=0.898$, $\epsilon_{i_{13/2}}=1.633$, 
$\epsilon_{f_{7/2}}=2.340$, and $\epsilon_{h_{9/2}}=3.414$.

\section*{Results}
In Fig. 1 we report the experimental \cite{Zhang96,Serg87}
and theoretical spectra of the
three-proton nucleus $^{135}$I. As regards the two-proton nucleus
$^{134}$Te, the results obtained by using the Bonn A potential are
to be found in Fig. 6. The spectra of both these nuclei have already
been presented in a previous paper \cite{Cov98}, where a detailed
comparison between theory and experiment is made. In that paper
we also reported the calculated $E2$ and $E3$ transition rates in $^{134}$Te
and compared them with the available experimental data. 
While we refer the reader for details to the above paper, we emphasize
here the very good agreement between the calculated spectra and
the experimental ones, as is shown by the value of the rms deviation
$\sigma$ \cite{sigma}, which are
106 keV and 58 keV for $^{134}$Te and $^{135}$I, respectively.
\begin{figure}[H] % fig 1
\centerline{\epsfig{file=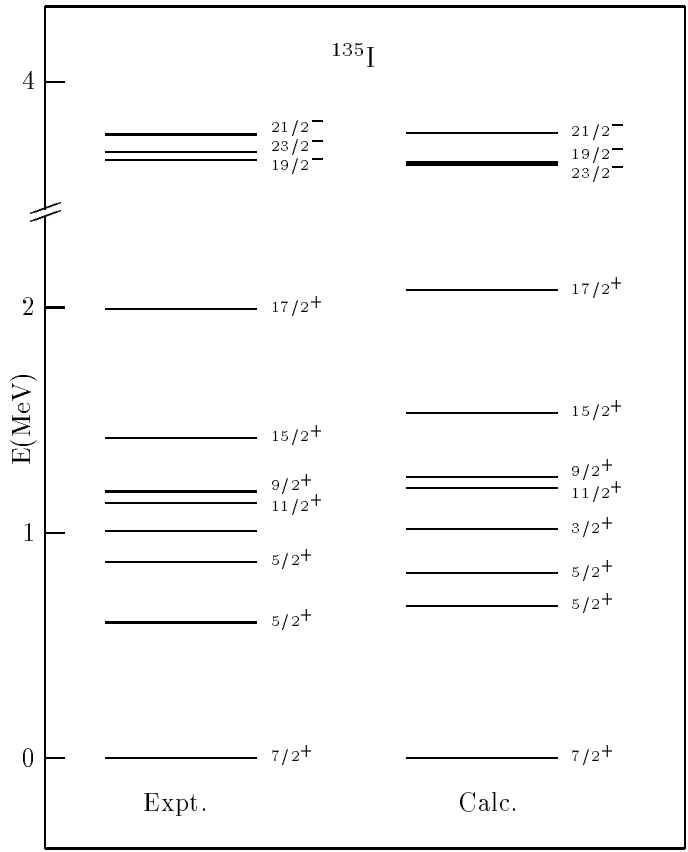}}
%\centerline{\epsfig{file=got.ps,height=3.5in,width=3.5in}}
\caption{Experimental and calculated spectrum of $^{135}$I.}
%\vspace*{10pt}
%\label{fig1}
\end{figure}
The experimental \cite{Tuli94} and calculated spectra of the four-proton
nucleus $^{136}$Xe are compared in 
Fig. 2, where all the calculated and experimental levels up to about  2.5 
MeV excitation energy are reported. Excluding the three states with 
$J^\pi=3^+,4^+$,  $J^\pi=(4^+)$, and $J^\pi=4^+$ at 2.13 , 
2.46, and 2.56 MeV, respectively, each level in the observed spectrum can 
be unambiguously identified with a level predicted by the theory.  The 
quantitative agreement between theory and experiment is quite 
satisfactory. In fact, a rather large discrepancy (248 keV) occurs only for 
the $0^+_2$ state, the calculated excitation energies of all other 
states differing by less than 110 keV from the experimental values. The rms 
deviation relative to the eight identified excited states is 107 keV. As 
regards the three above mentioned states, for which we have not attempted to 
establish a correspondence between theory and experiment, a firm spin 
assignment to the levels at 2.13  and 2.46 MeV is needed to clarify
the situation. 

Let us now come to the Pb isotopes. The experimental 
\cite{Helmer90,Rab93,Schmorak94} and theoretical
spectra of $^{206}$Pb, $^{205}$Pb and $^{204}$Pb are compared in Figs. 
3, 4 and 5, where we report all the calculated and experimental levels
up to 2.5, 1.5 and 2.0 MeV, respectively. In the high-energy regions
we only compare the calculated high-spin states with the observed
ones. From Figs. 3-5 we see that a very good agreement with experiment
is obtained for the low-energy spectra. In particular, in each of the
three nuclei the theoretical level density reproduces remarkably well
the experimental one.
\begin{figure}[H] % fig 1
\centerline{\epsfig{file=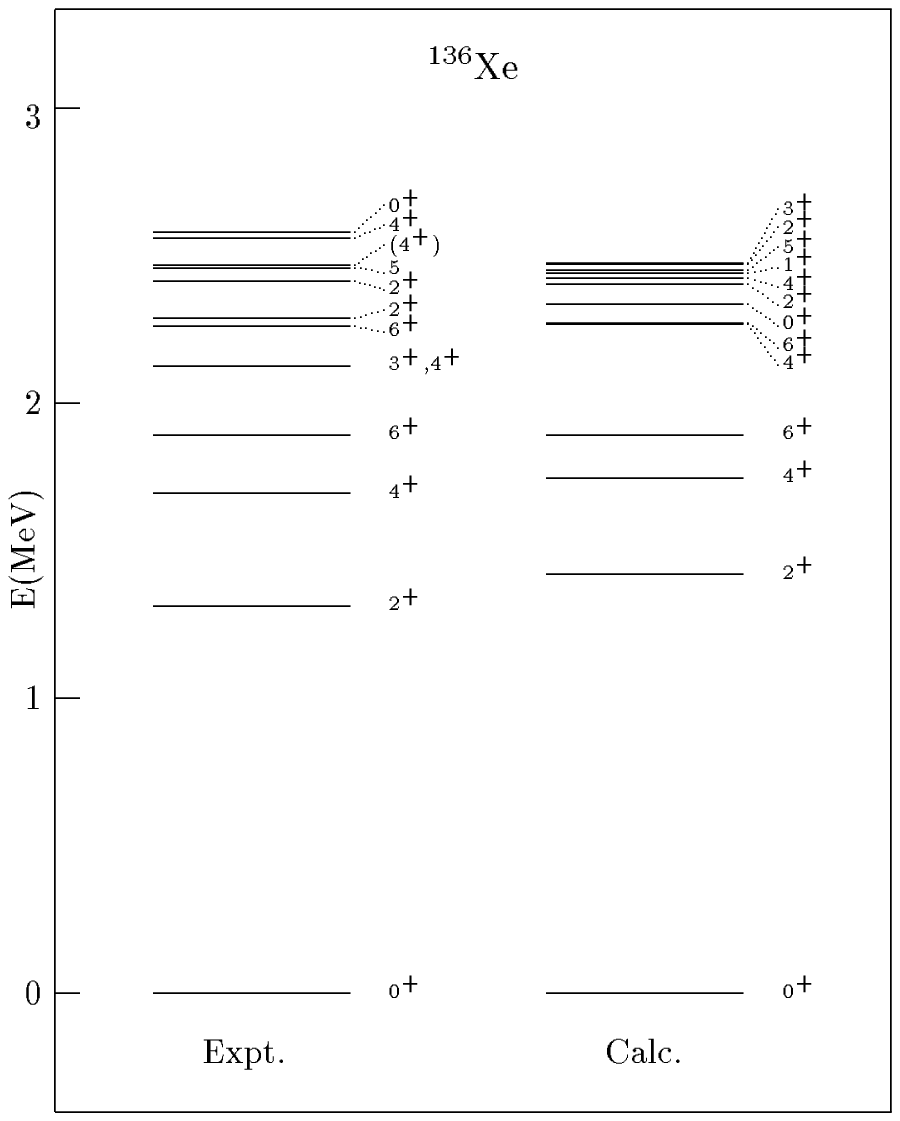}}
%\centerline{\epsfig{file=got.ps,height=3.5in,width=3.5in}}
\caption{Experimental and calculated spectrum of $^{136}$Xe.}
%\vspace*{10pt}
%\label{fig1}
\end{figure}
Note too that each state of a given $J^\pi$ in any of three calculated
spectra has its experimental counterpart, with a few exceptions. In fact,
as may be seen in Fig. 4, the ${5 \over 2}^-$, 
$({{3 \over 2}, {1 \over 2}})^-$, and $({{9 \over 2}, {7 \over 2}})^-$
states observed at 1.265, 1.374 
and 1.499 MeV in $^{205}$Pb cannot be safely identified with levels 
predicted by the theory. As regards $^{204}$Pb, we find the $0_4^+$
state at 1.954 MeV while the experimental one, which is not reported
in Fig. 5, lies at 2.433 MeV. It should be mentioned, however, that
the theory predicts four more $0^+$ states in the energy interval
2.2--2.6 MeV. Aside from these uncertainties, the agreement between
calculated and experimental spectra is such as to allow us to
identify experimental states with no firm or without spin-parity
assignment. For $^{206}$Pb our results suggest that the observed
levels at 2.197 and 2.236 MeV have $J^\pi = 3^+$ and $1^+$,
respectively. As for $^{205}$Pb, we predict $J^\pi$ =
${ {1 \over 2}^-}$  and ${ {3 \over 2}^-}$  for the experimental
levels at 0.803 an 0.998 MeV.

Regarding the quantitative agreement between our results and experiment,
the rms deviation $\sigma$ is 207 and 216 keV for
$^{206}$Pb and $^{204}$Pb, respectively. For $^{205}$Pb the $\sigma$ value
is 74 keV, excluding the three above mentioned states, for which we
have not attempted any identification.
Concerning the high-spin states in $^{206}$Pb and $^{205}$Pb, from
Figs. 3 and 4 we see that they are also well described by the
theory. In $^{204}$Pb the agreement between theory and experiment
is rather worse for the states lying above 4.3 MeV excitation energy,
the largest discrepancy being about 400 keV for the ${16}^+_2$ state. 
\begin{figure}[H] % fig 1
\centerline{\epsfig{file=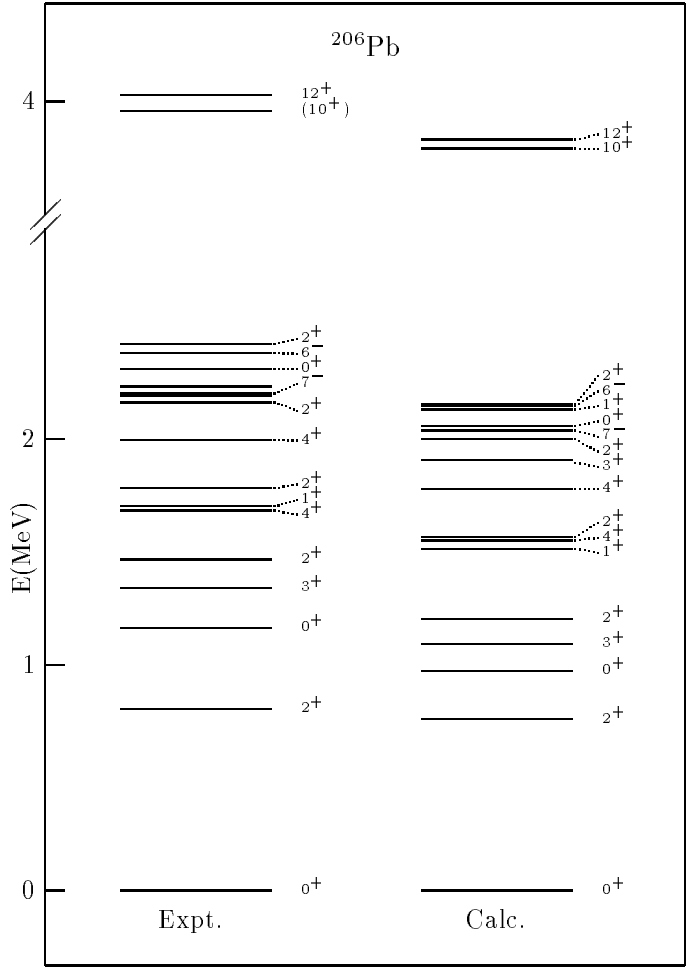}}
%\centerline{\epsfig{file=got.ps,height=3.5in,width=3.5in}}
\caption{Experimental and calculated spectrum of $^{206}$Pb.}
%\vspace*{10pt}
%\label{fig1}
\end{figure}
We have also calculated the electromagnetic properties for each of
the three isotopes. For the sake of brevity we do not report these
results in the present paper, but refer the reader to Ref. \cite{Cor98}, 
where a detailed comparison with the available experimental data 
is also made. We only mention here that a very good overall agreement
is obtained.

As already mentioned in the Introduction, we are currently
investigating the dependence of nuclear structure 
results
on the $NN$ potential used to derive the model space effective
interaction. Here we present the results obtained for the
nucleus $^{134}$Te which provides a very good testing ground
for this investigation. In fact, this nucleus has only two
valence protons and thus offers the opportunity to test
directly the matrix elements of the calculated effective
interactions. Furthermore, all the s.p. energies, except
$\epsilon_{s_{1/2}}$, are known from experiment. The uncertainty in
the position of the $s_{1 \over 2}$ level, however, has practically no influence on the 
low-energy spectrum. 
\begin{figure}[H] % fig 1
\centerline{\epsfig{file=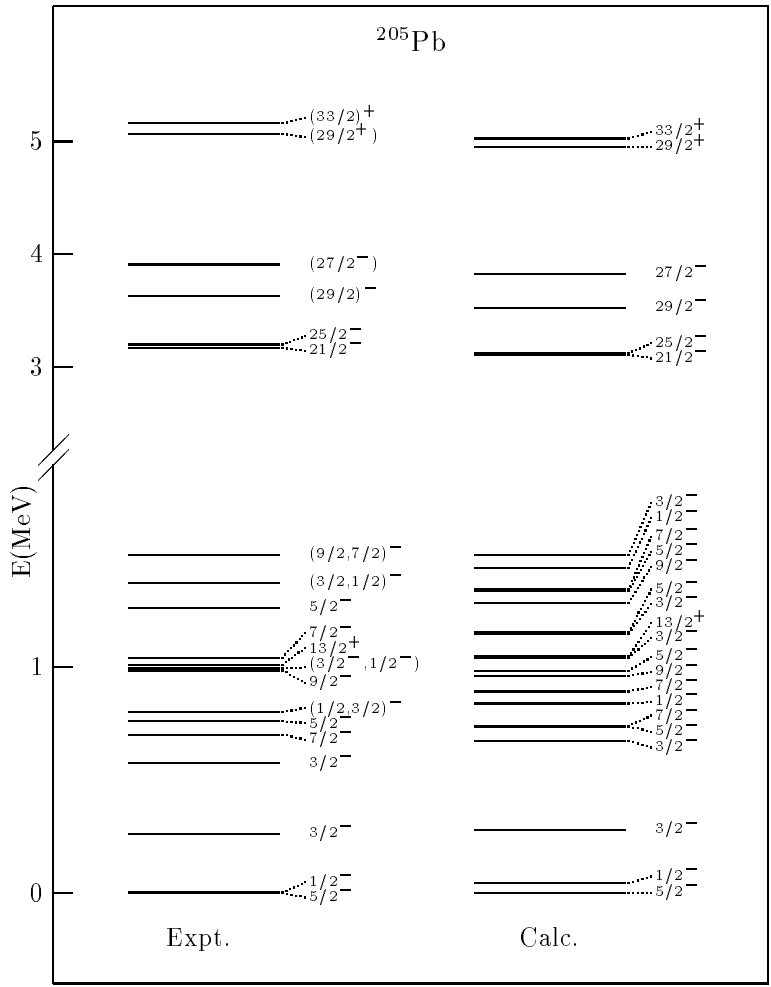}}
%\centerline{\epsfig{file=got.ps,height=3.5in,width=3.5in}}
\caption{Experimental and calculated spectrum of $^{205}$Pb.}
%\vspace*{10pt}
%\label{fig1}
\end{figure}

\begin{figure}[H] % fig 1
\centerline{\epsfig{file=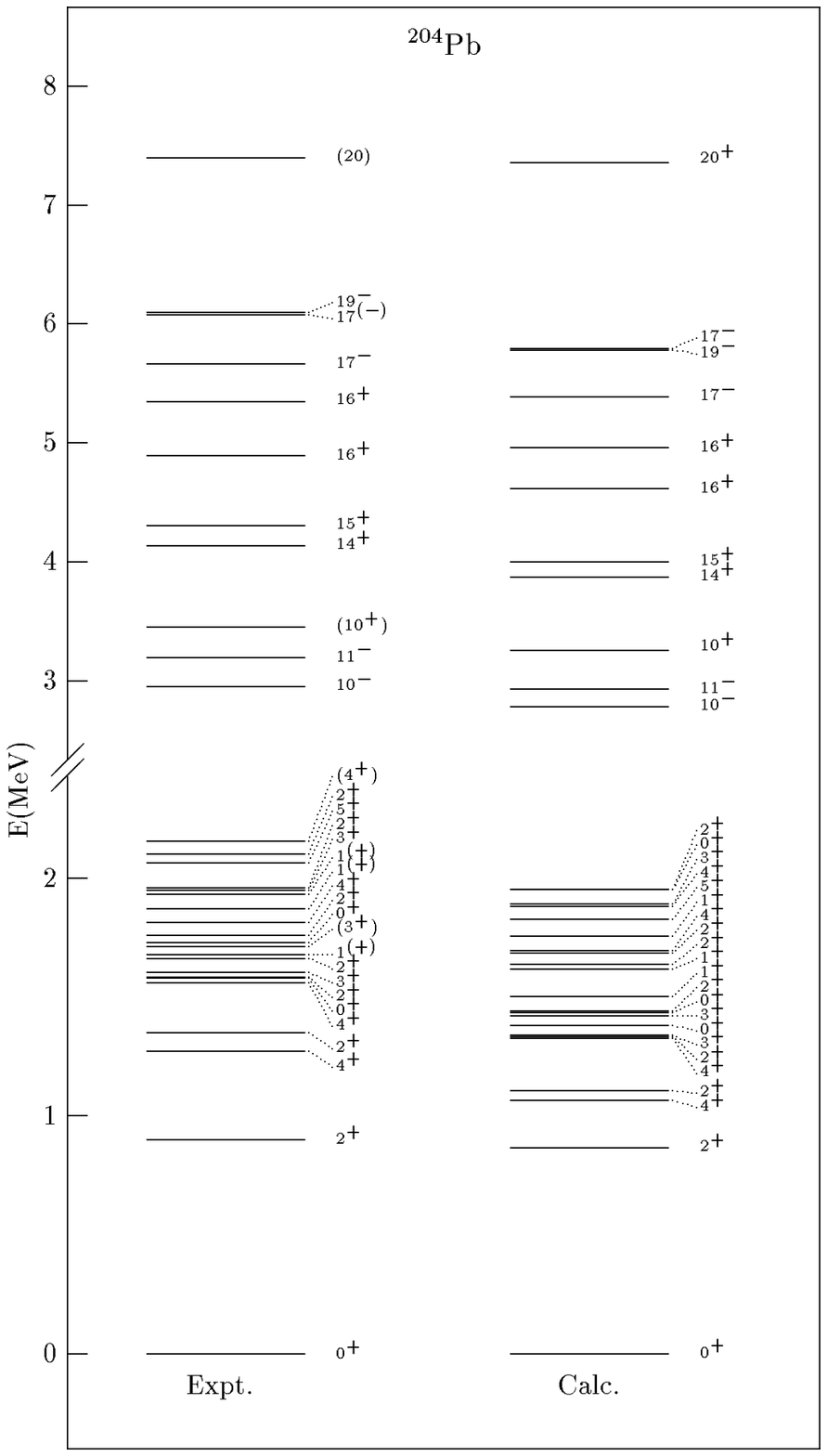}}
%\centerline{\epsfig{file=got.ps,height=3.5in,width=3.5in}}
\caption{Experimental and calculated spectrum of $^{204}$Pb.}
%\vspace*{10pt}
%\label{fig1}
\end{figure}
In Fig. 6 we report the four theoretical spectra
obtained by using the Paris, Nijmegen93, CD Bonn and Bonn A
potentials together with the experimental one \cite{Omt95,Serg94}.
We see that the best
agreement with experiment is produced by the Bonn A effective
interaction. The rms deviation $\sigma$ is 106, 160, 211,
and 346 keV for Bonn A, CD Bonn, Nijm93 and Paris,
respectively. These results show that different $NN$ potentials
produce somewhat
\begin{figure}[H] % fig 1
\centerline{\epsfig{file=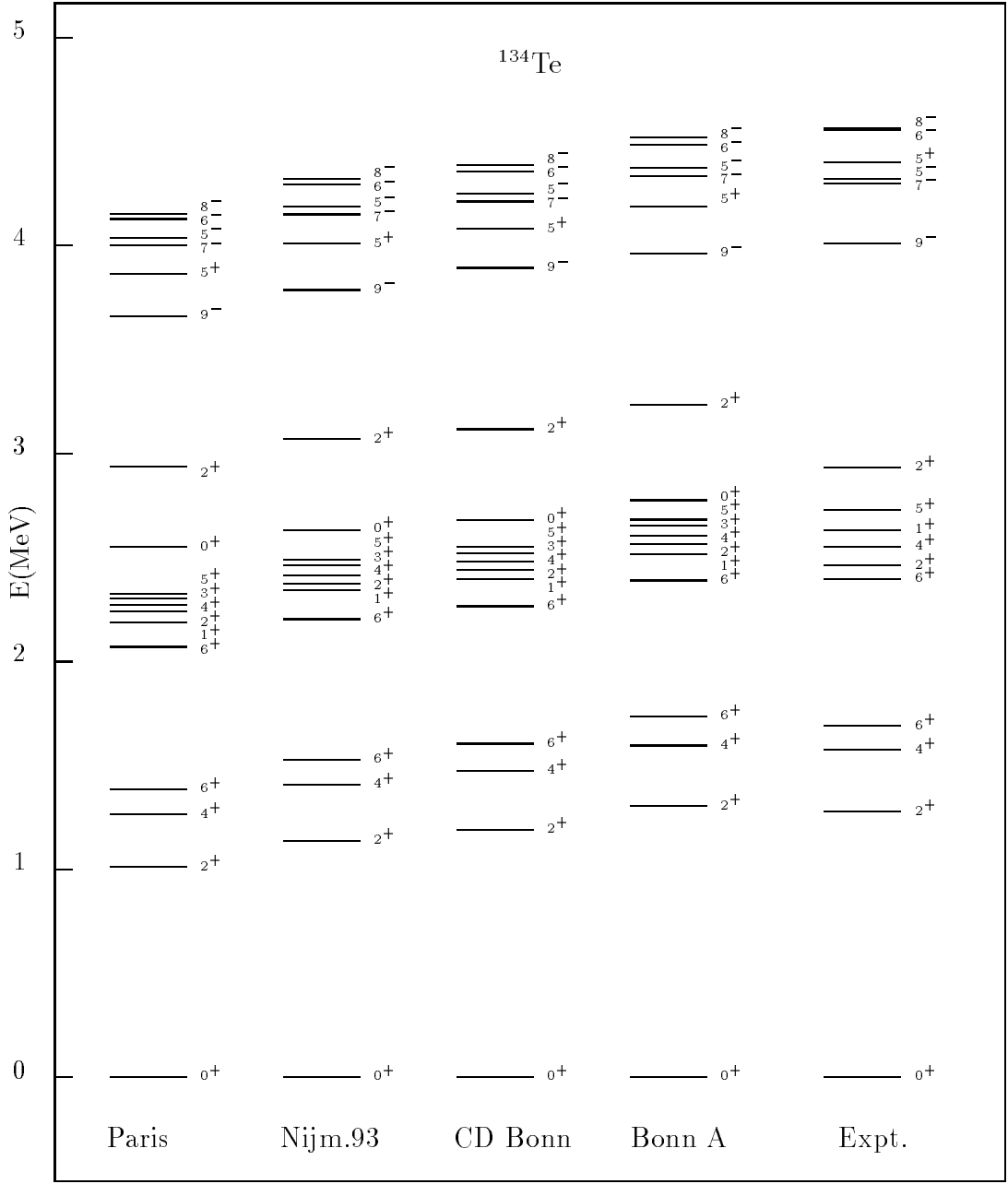}}
%\centerline{\epsfig{file=got.ps,height=3.5in,width=3.5in}}
\caption{Spectrum of $^{134}$Te. Predictions by various $NN$ potentials
are compared with experiment.}
%\vspace*{10pt}
%\label{fig1}
\end{figure}
\noindent
different nuclear structure results. Of course,
the significance of these differences, which do not exceed
100 keV if we exclude the Paris potential, may be a matter
of discussion. The following remark is, however, in order.
The four potentials considered differ in the strength of the
tensor-force component as measured by the predicted $D$-state
probability of the deuteron $P_D$. This is 4.4\% for Bonn A,
4.83\% for CD Bonn, 5.76\% for Nijm93, and 5.8\% for Paris.
Our results suggest that potentials with a weak tensor force
may lead to a better description of nuclear structure
properties. This is quite an interesting point since 
differences in $P_D$ may in turn be traced to off-shell differences. 

\section*{Closing remarks}
As pointed out in the Introduction, in this paper we have been concerned 
with the two main problems related to microscopic nuclear structure
calculations starting from a free $NN$ potential. On the one hand,
we have shown some recent results of shell-model calculations for 
nuclei around $^{132}$Sn and $^{208}$Pb obtained by employing an
effective interaction derived from the Bonn A nucleon-nucleon
potential. The success achieved by these calculations confirm
the conclusion reached in our earlier works \cite{Andr96,Andr97,Cov97,Cov98}
that this effective interaction is able to describe with quantitative
accuracy the spectroscopic properties of nuclei around closed
shells. 

On the other hand, we have presented some preliminary
results of a study aimed at ascertaining how much nuclear 
structure results depend on the $NN$ potential one starts with.  
As a first testing ground, we have chosen the two valence-proton
nucleus $^{134}$Te, making use of the Paris, Bonn A,
CD Bonn and Nijm93 $NN$ potentials. Since this study is still
in the initial stage, it is premature to draw any definite
conclusion. We plan to extend this kind of calculations to 
several other nuclei and to other modern high-quality potentials,
like Argonne $v_{18}$ \cite{Wir95}, Nijm I and Njim II \cite{Stoks94},
which have not been
considered here. The results obtained so far, however, 
indicate that microscopic nuclear structure calculations
may provide valuable information on the off-shell behavior of 
the $NN$ potential.
\section*{Acknowledgments}
This work was supported in part by the Italian Ministero dell'Universit\`a
e della Ricerca Scientifica e Te\-cnologica (MURST) and by the U.S. DOE
Grant No. DE-FG02-88ER40388. We would like to thank Ruprecht Machleidt
for providing us with the matrix elements of $NN$ potentials and for
valuable comments.


\begin{references}
\bibitem{sigma} We define $\sigma = \{ { {1 \over {N_d} }
\sum_i {[E_{\rm exp}(i) - E_{\rm calc}(i)]}^2 } \}^{1/2}  $, where $N_d$ 
is the number of data.
\end{references}
\end{document}